\documentclass{article}

\usepackage{arxiv}
\usepackage[utf8]{inputenc} 
\usepackage[T1]{fontenc}    
\usepackage{hyperref}       
\usepackage{url}            
\usepackage{booktabs}       
\usepackage{amsfonts}       
\usepackage{nicefrac}       
\usepackage{microtype}      
\usepackage{lipsum}         
\usepackage{graphicx}
\usepackage[
    backend=biber,
    style=ieee,
  ]{biblatex}
\usepackage{doi}
\usepackage{amsmath}
\usepackage[dvipsnames]{xcolor}

\title{Walking Through Complex Spatial Patterns of Climate and Conflict-Induced Displacements}

\newif\ifuniqueAffiliation

\ifuniqueAffiliation 
\author{ \href{https://orcid.org/0000-0000-0000-0000}{\includegraphics[scale=0.06]{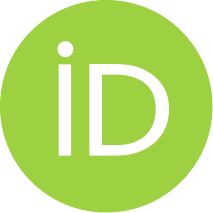}\hspace{1mm}David S.~Hippocampus}\thanks{Use footnote for providing further
		information about author (webpage, alternative
		address)---\emph{not} for acknowledging funding agencies.} \\
	Department of Computer Science\\
	Cranberry-Lemon University\\
	Pittsburgh, PA 15213 \\
	\texttt{hippo@cs.cranberry-lemon.edu} \\
	\And
	\href{https://orcid.org/0000-0000-0000-0000}{\includegraphics[scale=0.06]{orcid.pdf}\hspace{1mm}Elias D.~Striatum} \\
	Department of Electrical Engineering\\
	Mount-Sheikh University\\
	Santa Narimana, Levand \\
	\texttt{stariate@ee.mount-sheikh.edu} \\
}
\else

\usepackage{authblk}

\newbox{\orcid}\sbox{\orcid}{\includegraphics[scale=0.06]{orcid.pdf}}

\author[1]{%
	\href{}{\usebox{\orcid}\hspace{1mm} David Carranza\thanks{\texttt{dcarranzanav@uni-osnabrueck.de}}}%
}
\author[2]{%
	\href{https://orcid.org/0009-0000-4678-3301}{\usebox{\orcid}\hspace{1mm} Devansh Sharma\thanks{\texttt{d.sharma@ssmeridionale.it}}}%
}
\author[3]{%
	\href{https://orcid.org/0009-0002-7423-3491}{\usebox{\orcid}\hspace{1mm} Francisco Malveiro\thanks{\texttt{francisco.malveiro@isi.it}}}%
}
\author[4]{%
	\href{https://orcid.org/0000-0003-4903-0860}{\usebox{\orcid}\hspace{1mm} Gustavo Kohlrausch\thanks{\texttt{gustavo.luis.kohlrausch@gmail.com}}}%
}
\author[5]{%
	\href{https://orcid.org/0000-0002-4517-5328}{\usebox{\orcid}\hspace{1mm} Jisha Mariyam John\thanks{\texttt{jishamariyam.phd201010@iiitkottayam.ac.in}}}%
}
\author[6]{%
	\href{}{\usebox{\orcid}\hspace{1mm} Kaloyan Danovski\thanks{\texttt{kaloyan.danovski@upf.edu}}}%
}

\author[7]{%
	\href{https://orcid.org/0009-0005-4339-9925}{\usebox{\orcid}\hspace{1mm}Malvina Bozhidarova\thanks{\texttt{mbozhidarova@russell.co.uk}}}%
}
\author[8]{%
	\href{https://orcid.org/0000-0003-3167-1168}{\usebox{\orcid}\hspace{1mm}Rui Zheng\thanks{\texttt{rui.zheng@manchester.ac.uk}}}%
}
\author[9]{%
	\href{https://orcid.org/0000-0002-6198-7393}{\usebox{\orcid}\hspace{1mm}Sandro Sousa\thanks{\texttt{sfs@sodas.ku.dk}}}%
}

\affil[1]{Osnabruck University, Germany}
\affil[2]{Scuola Superiore Meridionale, Napoli, Italy}
\affil[3]{ISI Foundation, Turin, Italy}
\affil[4]{Physics Institute, Universidade Federal do Rio Grande do Sul, Brazil}
\affil[5]{Indian Institute of Information Technology Kottayam, India}
\affil[6]{Pompeu Fabra University, Barcelona, Spain}
\affil[7]{Russell Group Limited, Nottingham, United Kingdom}
\affil[8]{Alliance Manchester Business School, The University of Manchester, Manchester M15 6PB, United Kingdom}
\affil[9]{Center for Social Data Science (SODAS), University of Copenhagen, 1353 Copenhagen, Denmark}

\hypersetup{
pdftitle={Walking Through Complex Spatial Patterns of Climate and Conflict-Induced Displacements},
pdfsubject={physics.soc-ph, nlin.AO},
pdfauthor={David Carranza, Devansh Sharma, Francisco Malveiro, Gustavo Kohlrausch, Jisha
Mariyam John, Kaloyan Danovski, Malvina Bozhidarova, Rui Zheng, Sandro Sousa},
pdfkeywords={Migration, Internal displacement, Random walks, Diffusion on networks},
}

\begin{document}
\maketitle

\begin{abstract}
Extreme weather events are projected to intensify global migration, increase resource competition, and amplify socio-spatial phenomena, including intergroup conflicts, socioeconomic inequalities, and unplanned displacements, among others. Addressing these challenges requires consolidating heterogeneous data to identify, estimate, and predict the dynamical process behind climate-induced movements.
We propose a novel hybrid approach to reconstruct hazard-induced displacements by analysing the statistical properties of a diffusion process (walks) that explores the spatial network constructed from real displacements. The likely trajectories produced by the walks inform the typical journey of individuals, identifying potential hazards that may be encountered when fleeing high-risk areas. As a proof of concept, we apply this method to Somalia's detailed displacement tracking matrix, containing 20,220 movements dating from February 8 to June 18, 2025. We reconstruct the likely routes that displaced persons could have taken when fleeing areas affected by conflict or climate hazards. We find that individuals using the most likely paths based on current flows would experience mainly droughts and conflicts, while the latter becomes less prominent at every subsequent step. We also find that the probability of conflict and drought across all trajectories is widely dispersed, meaning that there is no typical exposure. This work provides an understanding of the mechanisms underlying displacement patterns and a framework for estimating future movements in areas expected to face increasing hazards. 

\end{abstract}


\section{Introduction}
\label{sec:intro}


Migration has been part of human history for ages. People move to find better lives, escape conflict, or seek new opportunities. Most migration is safe and organised, often driven by work and regional connections \cite{mcauliffe2024world}. However, an increasing number of movements have been driven by unplanned displacements resulting from armed conflict and natural disasters \cite{grid2025global}, which are correlated \cite{hsiang2011civil, koubi2019climate}. Furthermore, global heating has been increasing the frequency of catastrophes \cite{ospanova2024climate,dastagir2015modeling,stott2016climate}, which in turn affects mobility networks globaly. Estimating how distinct groups of people will move and what path they will take (migration flows) in the event of hazards is critical. The International Organization for Migration (IOM) estimates that improved data could be worth up to \$35 billion to governments and migrants \cite{kraly2020data,laczko2018more,chi2025measuring}.

A widely applied model to estimate migration, which takes its name from an analogy with Newton's law, is the gravity model \cite{ramos2016gravity}. In this model, the probability of interaction between two places is directly proportional to a size parameter, such as the population of a city, and inversely proportional to the distance.
Although widely adopted, the gravity model depends on adjustable parameters (such as population size exponents and distance decay factors) that can vary significantly across regions depending on the context. This makes the model less transferable and reduces its predictive power across different regions \cite{simini2012universal}. Additionally, the gravity model is static and provides a snapshot of migration flows at a given point in time. It does not account for how migration evolves or spreads over time in response to shocks, policy changes, or economic transitions \cite{beyer2022gravity}. 
Given the limitations of the gravity model, an alternative parameter-free approach, the radiation model, was introduced in \cite{simini2012universal}. However, a key drawback of the radiation model is its inability to accurately predict migration flows across all spatial scales, particularly for movements of very short or long distances. 

Diffusion on network models offers an alternative framework for understanding the temporal spread and dynamic evolution of migration \cite{rajulton1991migrability}.
The diffusion model treats the movement of people (or information, behaviours, etc.) as a spreading process, similar to how particles diffuse in physical systems or how diseases spread. In migration processes, diffusion models simulate how individuals or populations gradually relocate over time and space, influenced by environmental and social factors.
Building on the diffusion-based perspective, a random walk on a spatial network captures the step-by-step movement of individuals through a spatial system, reflecting the probabilistic nature of movements influenced by socioeconomic factors, geopolitics, and geographic accessibility \cite{noh2004random,fronczak2009biased,sousa2022quantifying}.
Random walk models have been widely applied across various domains to study human mobility. In the context of urban movement, they are used to capture the unpredictable and dynamic patterns of how individuals navigate within cities \cite{barbosa2018human}. Within transportation networks, random walks help model the movement of agents or vehicles through complex infrastructures, enabling the identification of critical routes and connections \cite{roubinet2022multi}. For epidemic spreading, these models simulate the spatial movement of individuals or pathogens, offering realistic insights into how infections propagate across regions \cite{bassolas2021diffusion, granger2025random, weiss1983random}. In the study of information diffusion, random walks are employed to model how information, ideas, or influence spread across social networks \cite{lam2012information,rosvall2008maps}.
In the context of migration, random walks provide a flexible framework for modelling how individuals or populations gradually relocate, accounting for multiple possible trajectories rather than assuming direct or optimal routes. 
For example, random walk models have been used to predict the potential destinations of internally displaced persons (IDPs) in conflict situations \cite{IOM2017NigeriaRandomWalk}. These models help capture the unpredictable and complex patterns of displacement, leading to more accurate forecasts of population movements and needs \cite{yasuda1975random}.

In this work, we focus on Somalia as a case study due to the country's struggles with several droughts, floods, and a multiple-year ongoing civil war \cite{ahmed2024nexus,thalheimer2023large}. Combined, the climate crisis and armed conflicts are rapidly increasing the number of IDPs in the region. Furthermore, the Somali economy relies heavily on agriculture, making this sector particularly vulnerable to extreme weather and potentially amplifying slow-onset displacements. The IOM collects and publishes displacement data for the region \cite{IOMSomaliaEmergencyTrend2025}. However, the data includes only the origin and destination of movements, lacking information on how people arrived at their destination and the level of danger encountered along the way. 

To gain insights into the paths and hazards faced by IDPs during their journey, we propose a diffusion-based model on a spatial network created using Somalia's displacement flows. The flows range from February 8 to June 18, $2025$. Since people move from one place to another, the network is directed, representing the origin of IDPs and their final destination. The links in the network are weighted by the volume of IDPs that moved from each of the connected regions. To examine the various hazards encountered along the possible paths of the network, we define the most common hazard that has occurred in each location as an attribute of the nodes. Then, we run a series of diffusion processes on the network, simulating potential difficulties encountered along every path.

We start by detailing the dataset and the methodology, followed by the results in Section \ref{sec:results}. We find that conflict and drought consistently appear as the most influential hazards, with drought risk being pervasive along nearly all migration trajectories. Additionally, exposure to hazards is not uniformly distributed but varies with path length and location. We conclude by discussing our findings, the limitations of our approach, as well as avenues for further research in Section \ref{sec:discussion}.

\section{Methods}
\label{sec:methods}
In the following, we outline the data and methodology. Section~\ref{s:data} provides a detailed description of the dataset. Section \ref{s:network_construction} describes the procedure for constructing a directed and weighted displacement network that illustrates displacement flows between settlements in Somalia. In Section \ref{s:diffuion_process}, we detail the setup of a diffusion process on the network. Lastly, Section \ref{s:simulation} explains the process of simulating the diffusion process on the network to assess how hazard-induced displacement affects spatial dynamics.

\subsection{Data}\label{s:data}
The Somalia Emergency Trend Tracking Dataset (February 2025)~\cite{IOMSomaliaEmergencyTrend2025} represented by a Displacement Tracking Matrix (DTM), is a crisis-focused dataset developed to monitor and analyse sudden displacement movements within Somalia. Since February 2025, DTM teams have collected data across 25 districts between February 8, 2025, and June 18, 2025. In the Somalia Emergency Trend Tracking Dataset, four levels of granularity refer to the hierarchical organisation of data based on administrative boundaries or spatial units. This structure allows the dataset to present information across multiple geographic scales, typically including country, region, district and settlement levels. We opt for the settlement level as it provides the most detailed granularity, indicating specific locations such as villages, neighbourhoods, or IDP camps where the data was collected.

The original dataset comprises 20,220 entries, each documenting an instance of displacement. To ensure relevance to the study's objectives, we retained variables including region and district names, dates of arrival and departure, cause of displacement, places of origin and destination, as well as demographic attributes such as gender and age group (male, female, children). In instances where the actual settlement was reported as the location of both interview and arrival and explicitly confirmed by respondents, we treated this as the final destination settlement.

A significant challenge in utilising the raw data stemmed from inconsistencies in settlement naming conventions. Settlement names were recorded in various ways, both in Somali and English, and frequently contained typographical errors, incomplete entries, or extraneous information. To address this, we undertook a systematic data cleaning process: origin settlement names were standardised by cross-referencing them with correctly spelt names from the current settlement registry. This matching was performed at the district and regional level, ensuring that only settlements within the same administrative district and region were considered equivalent. This process enhanced the geographic integrity and analytical reliability of the cleaned dataset.

Figure~\ref{fig:network}-A shows summary statistics on demographic composition, displacement history, and regions of origin. It reveals that the displaced population is predominantly composed of children under 18 years of age, who account for 43.65\% of all cases (101,620 individuals), followed by adult females (32.38\%, 75,386) and adult males (23.96\%, 55,788). This demographic profile highlights the particular vulnerability of children and women in the context of forced displacement and suggests an urgent need for targeted humanitarian responses.


\subsection{Network construction}
\label{s:network_construction}

We construct a spatial network based on movements between settlements. We define a graph $G = (V, E)$, where each node $v\in V$ corresponds to a settlement, and each edge $e \in E$ represents an observed displacement from one settlement to another during the analysed period. Edges between settlements $v_i$ and $v_j$ are directed from the origin to the destination settlement, reflecting the directionality of displacement movements. The weight of each edge is defined as the proportion of internally displaced persons (IDPs) departing from the source node ($s$) who arrive at the target node ($t$), relative to the total number of individuals who have left $s$ during the study period.

\begin{figure}[ht]
    \centering
    \includegraphics[width=1.0\linewidth]{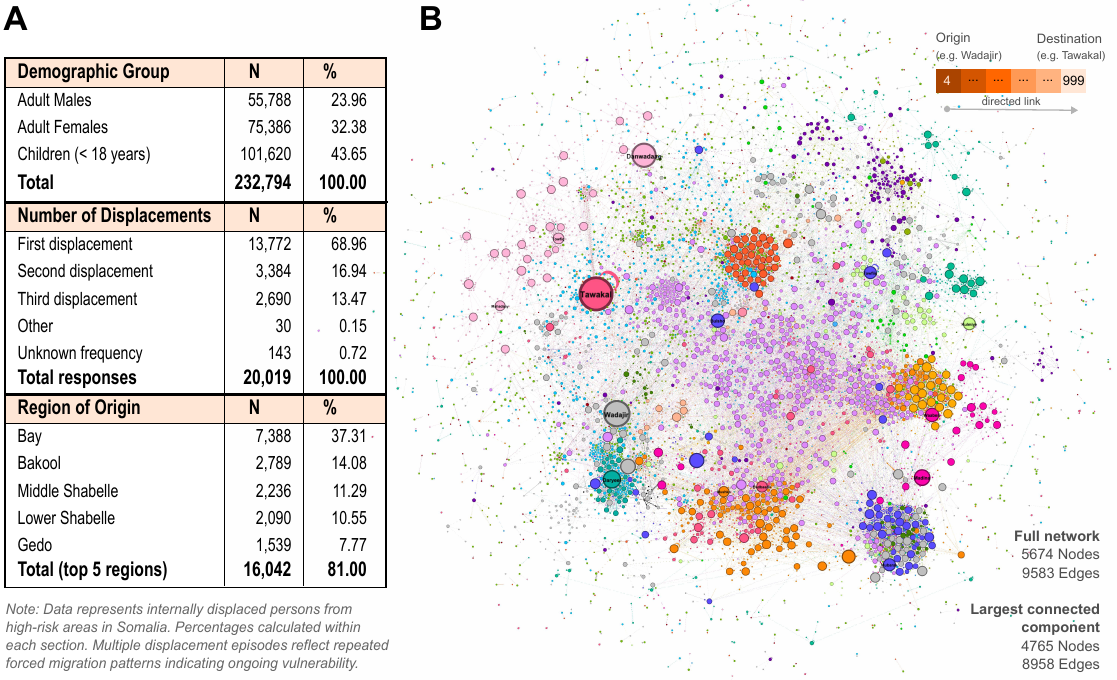}
    \caption{\small \textbf{Descriptive summary and constructed network of Somalia's displacement flows}. (A) Descriptive statistics of displacements, including demographic group, number of displacements, and top regions of origin. The majority of movements recorded in the dataset are associated with the first displacement of the population. Children are the most represented demographic group across all displacements. (B) Graphical representation of the constructed network. Node sizes are proportional to their in-degree. Colours represent membership in distinct districts of Somalia. For visualisation purposes, we only colour the $20$ districts with the largest flow of IDPs.}
    \label{fig:network}
\end{figure}

Applying this methodology to the \cite{IOMSomaliaEmergencyTrend2025} dataset results in a network comprising 5,674 nodes and 9,583 edges (Figure~\ref{fig:network}-B). Notably, this approach produces a considerable number of node pairs that are isolated from the main network. Since simulations starting from any of those isolated nodes would be constrained within disconnected components, we restrict our simulation analysis to the largest connected component of the network.

The largest connected component has 4,765 settlements, interconnected by 8,958 edges, indicating a densely linked core of the displacement system. Notably, several settlements, such as \textit{Tawakal} and \textit{Banaadir}, emerge as prominent hubs, reflecting their centrality within the displacement pathways. 

Formally, we can denote the probability of a displacement from node \( i \) to node \( j \) as \( P(i \to j) \). We then consider $P(i \to j) \propto w_{ij}$ where \( w_{ij} \) is the total number of IDPs that migrated from node \( i \) to node \( j \).
For each source node \( i \), we normalise the weight of each outgoing edge by the total number of individuals leaving settlement \( i \). Thus, we have 
\begin{equation}\label{eq:prob}
   P(i \to j) = \frac{w_{ij}}{\sum\limits_{j\in V} w_{ij} A_{ij}}, 
\end{equation} where $A$ is the adjacency matrix of the constructed network, disregarding the weight of the edges. 

\subsection{Diffusion processes on networks}  
\label{s:diffuion_process}
    
 A diffusion process on a graph $G$ is a discrete-time stochastic process that starts from a given node $v\in V$ and proceeds through a sequence of nodes ${v_0, v_1, v_2......v_k\in V}$ corresponding to the times ${t_0, t_1, t_2,......t_k}$, with a well-defined transition probability rule (\cite{newman2018networks}). In an unbiased diffusion process, for each node, there is a uniform probability distribution of moving to any of its neighbours. Conversely, a biased process assigns a higher probability of moving to a specific set of nodes based on a given criterion, such as node degree, edge weights, or another node attribute. This bias allows a walk on the network to move preferentially towards more attractive nodes. The biased diffusion processes are particularly useful for modelling realistic behaviours such as displacement towards more economically appealing countries or homophily-driven segregation \cite{prietocuriel2024diaspora}.

In displacement flow studies, a random walk from a source to a target node represents the displacement trajectory taken by individuals or groups as they relocate from one place to another, influenced by factors such as socioeconomic preferences, policy restrictions or environmental conditions.

In a uniform walk, the agent chooses among all available out-neighbours with equal probability. The transition probability is given by 
\begin{equation}\label{eq:unbia}
    P(i \to j) =\frac{1}{\sum\limits_{j\in V}A_{ij}}.
\end{equation} This corresponds to an unbiased diffusion process, where movement is governed solely by the graph's connectivity.
In a biased walk, the transition probability depends on edge weights. Here, we consider the probability that the biased random walk follows the weight of the edges, i.e., following Eq.\eqref{eq:prob}.

\begin{figure}[ht]
    \centering
    \includegraphics[width=0.8\linewidth]{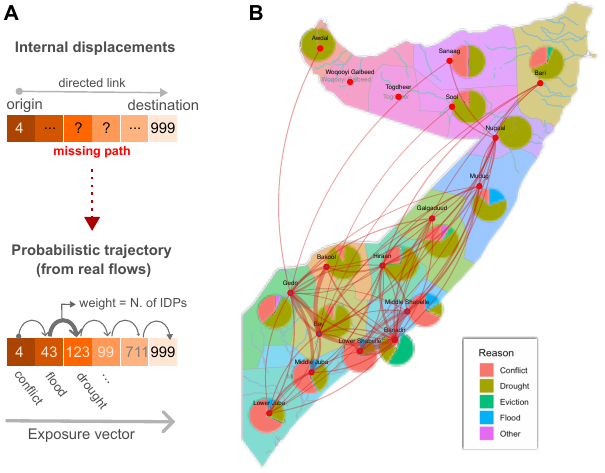}
    \caption{\small \textbf{Path reconstruction and spatial displacement flows}. (A) Reconstruction of the path from an origin to a destination point using the displacement matrix. Starting with an origin-destination pair (top panel), running simulations allows us to estimate a potential path between those points (bottom panel). Numbers in the figure correspond to node ID. Such paths enable us to identify the hazards people are likely to encounter if they undertake that journey. (B) Representation of the most probable hazard to occur in each region of Somalia and displacements between regions. The width of the links represents the number of IDPs who took that journey. The southern region exhibits the greatest density of movements, and it is also a region with a high incidence of conflicts.}
    \label{fig:flows}
\end{figure}

\subsection{Simulation procedure}
\label{s:simulation}

To quantify how hazard-induced displacement influences spatial dynamics across the network, we model the spread of the impact using probabilistic paths (Fig.~\ref{fig:flows}-A). Each node (settlement) in the network is annotated with the number of people who relocated as a result of a specific hazard event. For each ordered pair of nodes $(s,t)$, we simulate $M=1000$ probabilistic paths that begin at the source node $s$ and aim to reach a target node $t$ on the network described in Section~\ref{s:network_construction}. Each probabilistic path proceeds one step at a time along the network's edges. A path is terminated under one of the following conditions: it reaches the target node $t$, it arrives at a sink node, defined as a node with out-degree zero, or it exceeds a maximum of $1000$ steps without reaching the desired target. These termination rules ensure computational efficiency and prevent probabilistic paths from running indefinitely or failing to achieve adequate spatial coverage.

The overall exposure of a probabilistic path to a hazard is computed as the proportion of affected individuals across the sequence of visited nodes. In more detail, each node \( v \in V \) in the network is associated with a set of hazard-specific displacement counts. For a given hazard \( h \), we denote the number of individuals displaced from node \( v \) due to the hazard \( h \) as \( D_h(v) \). The total displacement from node \( v \), across all hazards, is given by
\begin{equation}
    D_{\text{total}}(v) = \sum_{h \in \mathcal{H}} D_h(v),
\end{equation}
where \( \mathcal{H} \) is the set of all considered hazard types.

Now, consider a path \( \mathcal{P} = (v_1, v_2, \ldots, v_k) \) through the network. The total displacement along the path for hazard \( h \) is computed as
\begin{equation}
D_h(\mathcal{P}) = \sum_{i=1}^k D_h(v_{i}).
\end{equation}
The \emph{exposure probability} for the path \( \mathcal{P} \) to hazard \( h \) is defined as
\begin{equation}
E_h(\mathcal{P}) = \frac{D_h(\mathcal{P})}{D_{\text{total}}(\mathcal{P})}.
\end{equation}
The quantity \( E_h(\mathcal{P}) \in [0,1] \) represents the fraction of displacement along the path attributed to the hazard \( h \), and can be used to quantify the hazard-specific exposure for a path. Aggregating these exposure profiles across all simulated paths provides a probabilistic estimate of the hazard exposure in the network.

Understanding which types of hazards are the most influential in shaping displacement trajectories requires analysing the nature of movement in the early stages of relocation. We focus on estimating the likelihood of encountering different hazard types during the initial steps of simulated probabilistic walks across the mobility network. By identifying the IDP-weighted frequency of each hazard at each visited node during the first $k$ steps of each walk, we capture how specific hazards imprint on early movement behaviour. We show an example with $k=5$ steps in Figure \ref{fig:hazards}B). Aggregating this information across multiple walks initiated from every node, we construct a global profile of hazard prevalence during the onset of displacement. This provides insights into which hazard types are most likely to initiate population movement and drive local decision-making in the early phases of spatial diffusion.

Using the simulation procedure described previously, for each ordered pair of nodes \( (s, g) \in V \times V \), we generate \( M = 1000 \) random walks:

\[
\mathcal{P}^{(j)}_{s \to g} = \left(v^{(j)}_1, v^{(j)}_2, \dots, v^{(j)}_{m_j}\right),
\]

where \( j = 1, \dots, M \), and \( m_j \) is the number of steps in the \( j \)-th walk. We then consider only the first \( k = 5 \) steps of each walk (or fewer, in case the walk terminates earlier) when we analyse the early-step behaviour shown in Figure \ref{fig:hazards}B). Note that for the other analysis, we examine all the steps taken by the walker.


For a fixed source node \( s \in V \), we consider all random walks \( \mathcal{P}^{(j)}_{s \to t} \) that originate from \( s \) and define a weighted frequency for each hazard:

\begin{equation}
    P_i(h \mid s) = \frac{1}{M_s^{(i)}} \sum_{g \in V} \sum_{j=1}^{M} \mathbb{I}[i \leq m_j] \cdot D_h(v^{(j)}_i),
\end{equation}

where \( M_s^{(i)} \) is the total number of walks from \( s \) that reached at least step \( i \) and  \( \mathbb{I}[\cdot] \) denotes the indicator function. This results in a distribution \( P_i(h \mid s) \) for each source node \( s \) and each step \( i \), representing the likelihood of encountering a hazard \( h \) at that point in displacement paths starting from \( s \).

Finally, we average the per-node distributions over all source nodes \( s \in V \) to obtain the overall hazard likelihood at step \( i \):

\begin{equation}
    \bar{P}_i(h) = \frac{1}{|V|} \sum_{s \in V} P_i(h \mid s).
\end{equation}
This gives us, for each step \( i \in \{1, \dots, k\} \), a global distribution \( \bar{P}_i(h) \) representing the likelihood of encountering each hazard type \( h \in \mathcal{H} \) during the early phases of displacement paths across the entire network.

\section{Results}
\label{sec:results}

The analysis of displacement flows between settlements reveals substantial heterogeneity in both the scale and directionality of population movements across Somalia during the study period (February–June 2025). The majority of displacement flows are concentrated among settlements in the southern and central regions, with particularly high volumes observed between settlements in Bay, Banadir, Bakool, Lower Shabelle, and Gedo (see Figure~\ref{fig:flows}-B). These corridors reflect both the major routes and destination points for internally displaced persons (IDPs), highlighting spatial patterns of recurrent and large-scale displacement.

A close examination of the map reveals that some settlements, such as those in Banadir and Lower Shabelle, serve as principal receivers of displaced populations, while others in Bay and Bakool act as major origins or transit hubs. This asymmetric flow structure underscores the presence of persistent "push" and "pull" factors driving the movement of IDPs across the country.

The primary reasons for the displacement, as reported by respondents, are overwhelmingly linked to acute and compounding threats. Conflict and insecurity remain the dominant causes, accounting for a significant share of all recorded displacement episodes. In addition, climate-related shocks—notably drought and flooding—are frequently cited as key drivers, either independently or in combination with conflict. In many cases, these crises interact, forcing individuals and families to move multiple times in search of safety and basic necessities. More detailed information on displacement at the district level is shown in Figure~\ref{fig:reasons_displace}.


\begin{figure}[ht]
    \centering
    \includegraphics[width=1\linewidth]{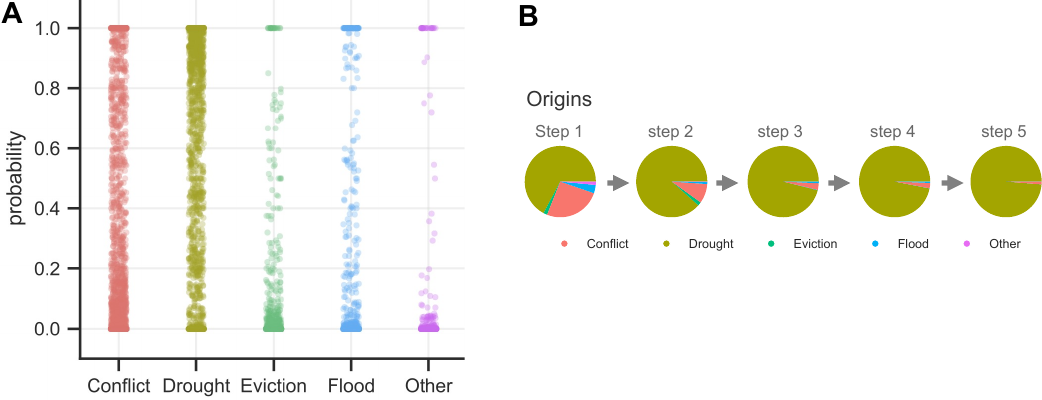}
    \caption{ \small \textbf{Hazard probabilities and their time dependency}. (A) Density of probabilities of hazard exposure for paths starting from distinct settlements. Each dot in the scatterplot represents the probability of encountering the respective hazard over all simulated trajectories, starting from a single location. Conflict and drought display a wider dispersion of probability of occurrence across the path compared to the other hazards. (B) Proportion of hazard exposures as a function of the number of steps taken from the origin location, up to a maximum of $5$ steps. Each pie chart represents the probability of exposure $k$ steps after the origin location. Probabilities are calculated based on all paths starting from an origin location and over all locations as origins. Drought becomes more predominant than the other hazards across the first five steps.}
    \label{fig:hazards}
\end{figure}

We can gain a deeper insight into displacement patterns by studying highly aggregated descriptions of the exposure vectors collected through our numerical experiments. Figure~\ref{fig:hazards} shows the exposure probabilities for paths starting from fixed sources. Fig.~\ref{fig:hazards}-A displays the expected density for encountering different hazard types along typical probabilistic trajectories. We weigh each exposure by the number of IDPs that left the location due to the given hazard at any point in the period represented in our data.

First, we note that these results confirm our primary observation that droughts and conflicts are predominantly responsible for driving the displacement of individuals. Drought risk, in particular, is a highly probable hazard along any path, regardless of its length, starting from any location. These two main hazards exhibit a higher average probability of exposure compared to other hazards, and their exposure distributions are more dispersed, implying lower predictability of exposure along typical paths. The exposure probability distribution for other hazards is significantly more skewed. Most locations show a negligible propensity for IDPs to encounter evictions, floods, and other extreme events on their path to final relocation.

Interestingly, extreme values are present in all hazard exposure probabilities. 
Since most source-destination paths involve only two or three locations (see Supplementary Figure~\ref{fig:path_lengths_dist}), these local displacements affect the statistics for exposure probabilities. This highlights the need to consider the distribution of exposure probabilities as a function of the number of movements along the path, shown in Fig.~\ref{fig:hazards}-B. There, we observe that the distribution of hazard exposure is not independent of the number of steps from the location of origin. Droughts are the most likely risk factor at all steps along the paths, but other hazards are still significantly represented within a few steps from the origin. This suggests that IDPs tend to be repelled by life-threatening and highly localised hazards such as conflicts and floods more strongly than by ubiquitous hazards such as droughts. Furthermore, the rate of decay of exposure probabilities depends on the type of hazard. For example, the risk of conflicts decreases more gradually than that of floods.



\section{Discussion}\label{sec:discussion}

This work introduces a diffusion-based approach to modelling population displacement patterns in response to conflict and climate hazards, using a spatial network of settlements annotated with hazard-specific relocation counts. By simulating probabilistic paths between source-target node pairs and analysing hazard exposure along those paths, we gain insights into the likelihood and timing of hazards that influence displacement trajectories. Our main contributions are methodological, and we use Somalia's IDP data as a proof of concept.

Nevertheless, our results highlight two important aspects of hazard-driven migration flow. On one hand, we have shown that there is a heterogeneous spatial distribution of hazards that affects the pattern of displacements. On the other hand, path-dependent exposure risk shows a separation of hazards based on the likelihood and timescale of risk exposure.

From Figure \ref{fig:hazards}-B, we learned that droughts are the predominant risk factors at all stages of displacement, while the exposure risk to other hazards decays along paths. Quickly developing hazards (such as floods) promote a quick avoidance path, even if this makes it more likely to encounter a different hazard (such as a drought). However, this rate of avoidance could also be explained by the fact that highly localised events can be escaped in fewer steps.

Unlike previous studies that primarily analyse displacement in static terms or focus on single drivers such as conflict or drought (\cite{ghosh2022systematic,zhou2024environmental,albarosa2022forced}), our work introduces a path-dependent, multi-hazard framework that quantifies how different types of risks shape displacement trajectories. By linking exposure probabilities to specific hazards along IDP pathways, we move beyond aggregate counts and static correlations, offering a dynamic and granular view of displacement behaviour under compound crises.

The weighted simulation framework incorporates observed displacement volumes as edge weights, allowing the walk process to reflect empirically grounded route preferences. The use of multiple simulated paths for each source-target pair also captures the inherent uncertainty and variability in individual displacements, acknowledging that not all individuals follow a typical route. Moreover, by aggregating simulated paths across the network, our method reveals dominant corridors and transit hubs that emerge from local interactions—insights that would be missed in purely deterministic or spatially localised models.

It should be noted that, despite the trajectory estimates presented here being based on real displacement movements, they neglect detailed information about the actual journeys taken by IDPs. For instance, a path from Middle Juba to Bari might have been taken directly (by flight) or included multiple transfers across settlements (by land transport). We assume that trips passing by settlements between these two regions are a viable route, but the inclusion of local mobility data and transport modes would inform this assumption.
We do not directly predict the path taken by individuals but estimate a path that they could have taken.

More generally, the non-trivial distribution of temporal and spatial scales for the development of different hazards likely introduces a set of pressures shaping the flow of displacement that is difficult to characterise and predict. 
However, the analysis is limited by the granularity, scope and quality of input data. The presence of inconsistent spellings requires manual or algorithmic name normalisation. Misclassification or missed duplicates could distort network topology and create ambiguities in the analysis. The dataset only includes observed movements between certain node pairs. Additionally, many low-volume or informal displacement routes may be overlooked, resulting in an incomplete representation of the entire displacement network. The simulation treats each probabilistic path as independent, whereas in reality, migration patterns are influenced by previous movements, social networks, and path dependence — none of which are captured in the data.

Additionally, the current approach does not incorporate temporal dynamics, i.e., when hazards occurred or how long displacements lasted. Hence, a potential direction for future research involves using temporal networks where edge weights or node attributes change over time. In addition, incorporating transit data would make trajectory estimates more realistic and practical for IDPs and policymakers organising evacuations. Despite these limitations, the approach proposed here provides a promising, scalable and generalisable framework for modelling hazard-induced displacement.


\section*{Acknowledgements}
This work is the output of the Complexity72h workshop, held at the Universidad Carlos III de Madrid in Leganés, Spain, 23-27 June 2025. https://www.complexity72h.com

\printbibliography


\clearpage

\renewcommand{\thefigure}{S\arabic{figure}}
\renewcommand{\thetable}{S\arabic{table}}
\renewcommand{\thepage}{\arabic{page}}
\renewcommand{\thesection}{S\arabic{section}}

\setcounter{figure}{0}
\setcounter{table}{0}
\setcounter{section}{0}
\setcounter{subsection}{0}

\vspace{.5cm}

\section{\label{sec:variables} Supplementary Figures}

\begin{figure}[ht]
    \centering
    \includegraphics[width=1\linewidth]{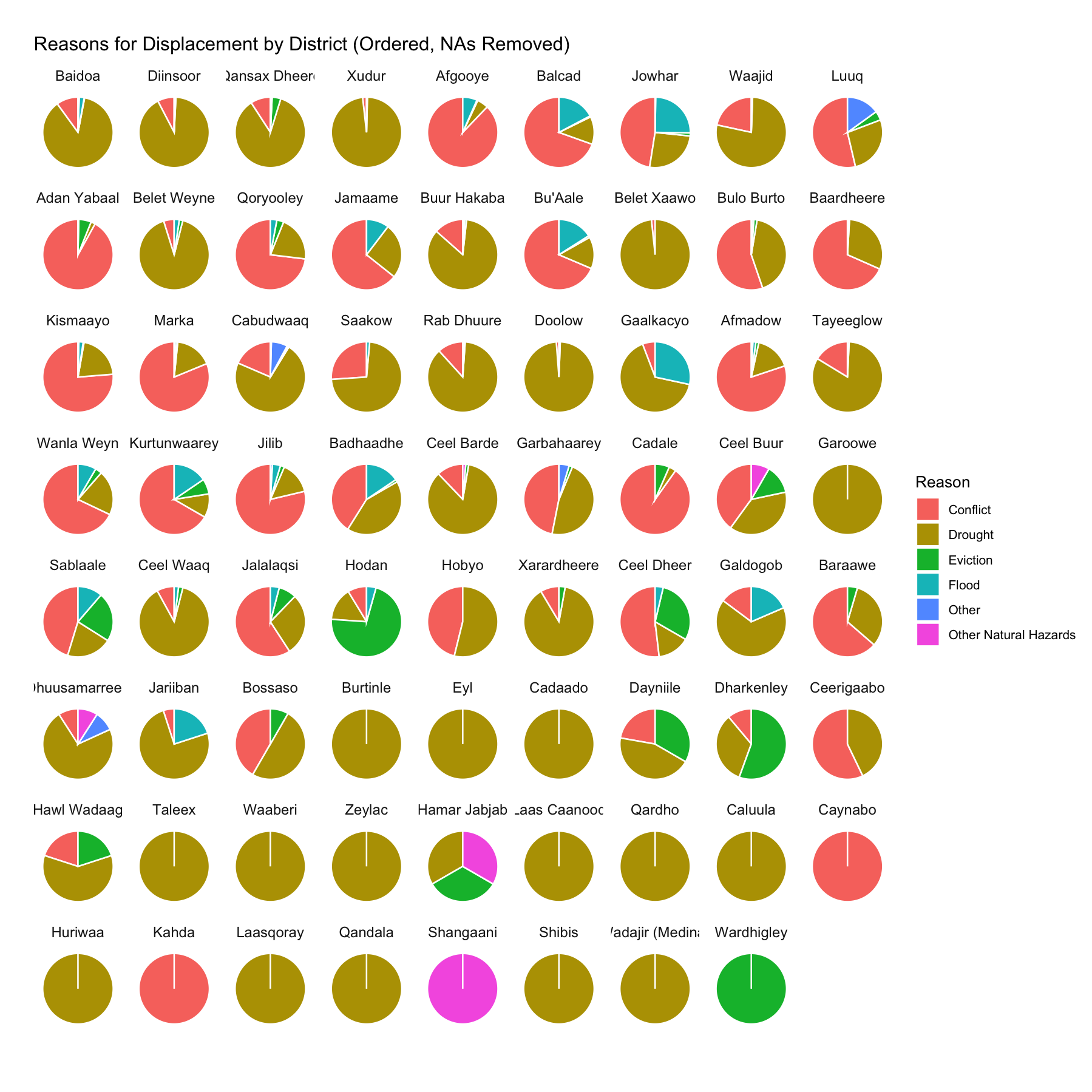}
    \caption{Reasons of displacement for $71$ districts in Somalia. Districts are presented in decreasing order of number of recorded displacements. Drought and conflict are predominant reasons for displacement across most regions. }
    \label{fig:reasons_displace}
\end{figure}

\begin{figure}[ht]
    \centering
    \includegraphics[width=0.7\linewidth]{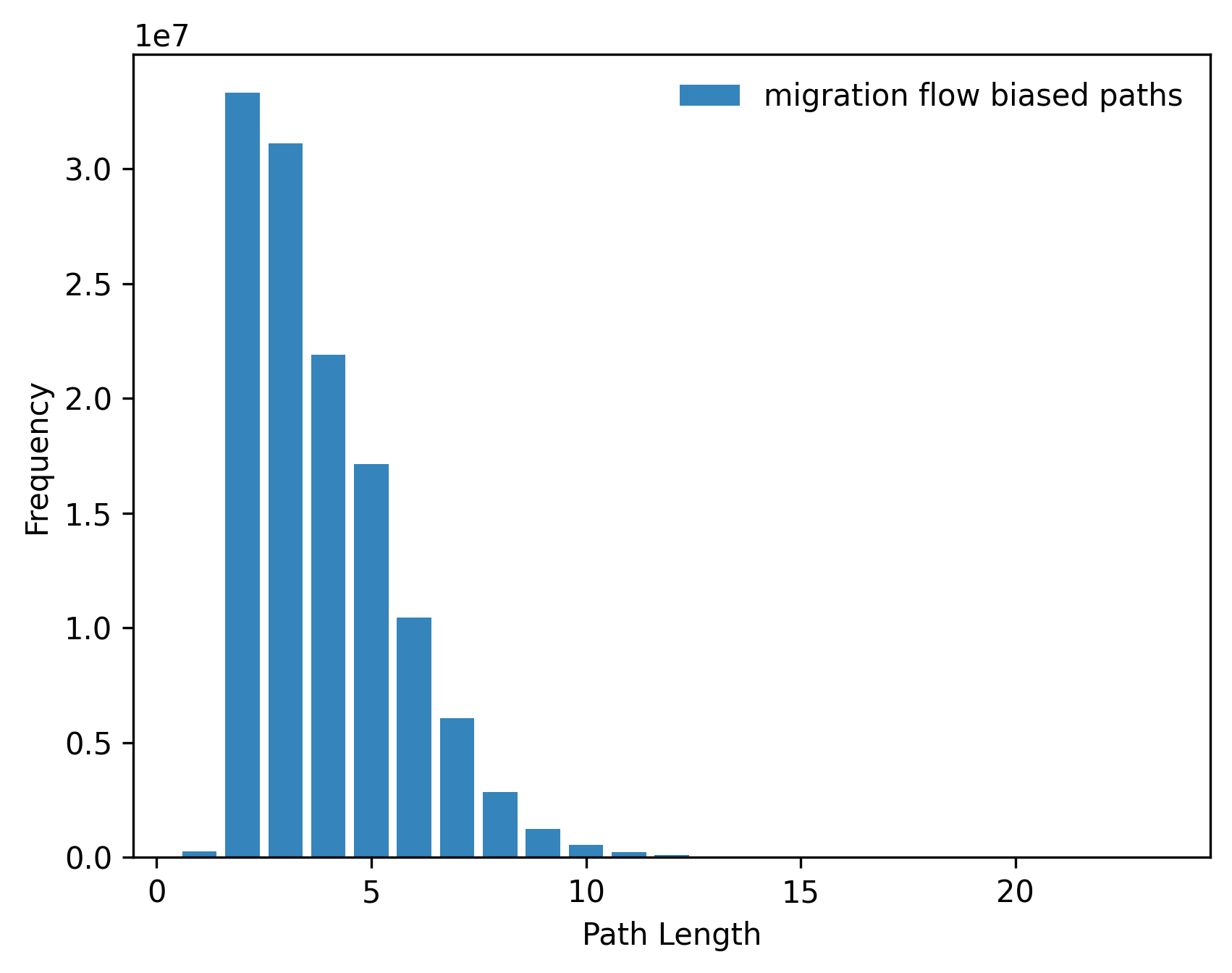}
    \caption{Frequency of path lengths observed in numerical simulations of random walks, aggregated over $M$ simulated trajectories for each of the $N$ locations as origin. Most paths are of short length, with a maximum length of 24 steps.}
    \label{fig:path_lengths_dist}
\end{figure}

\end{document}